\documentclass[epjCONF]{svjour}
\usepackage{graphics}
\usepackage[varg]{txfonts} 
\usepackage[latin1]{inputenc}

\session-title{A Universe of dwarf galaxies - Observations, Theories, Simulations}

\begin{document}

\title{UCDs as Probes of the Major and Minor Merger Histories of Galaxies}
\author{Mark A. Norris\inst{1}\fnmsep\thanks{\email{manorris@physics.unc.edu}} \and Sheila J. Kannappan\inst{1}}

\institute{$^{1}$~Dept. of Physics and Astronomy, University of North Carolina, Chapel Hill}

\abstract{
Two competing theories posit that Ultra Compact Dwarfs (UCDs) form either as the
stripped nuclei of dwarf galaxies or as giant globular clusters (GGCs) associated
with the largest globular cluster (GC) systems. By focussing on the field and group environments where 
young UCDs may be most common, we have discovered the first UCD that is clearly 
the result of recent ($<$4 Gyr ago) stripping of a companion galaxy. However, we have
also found a definitive case of a multiple-UCD system created via GC formation 
processes, which are likely associated with major galaxy mergers. We demonstrate that 
it is possible to reliably distinguish the two types of UCD, thereby probing 
both the major and minor merger histories of individual galaxies.} 

\maketitle

\section{Introduction}
\label{intro}
Ultra Compact Dwarfs are stellar systems with radii and masses 
intermediate between those of globular clusters and dwarf galaxies. UCDs 
have mainly been studied in galaxy clusters but have recently been noted in 
non-cluster environments as well (e.g. \cite{Hau09}). Intrigued by the possibility 
that UCDs are common in all environments, we undertook a search for UCDs 
associated with field/group galaxies. We visually examined archival HST WFPC2 
or ACS imaging of 76 galaxies visible from the southern hemisphere 
during spring, discovering 11 candidate UCDs associated with 9 individual galaxies. 
In these proceedings we focus on three systems with spectroscopically confirmed 
UCDs: NGC3923 (2 newly confirmed + 1 probable), NGC4546 (1 newly confirmed), 
and the Sombrero (1 UCD discovered by \cite{Hau09}). Using new SOAR 
spectroscopy, archival SAURON spectroscopy, and HST imaging we find evidence for 
two distinct UCD formation channels. One is associated with GC formation, producing
GGCs that follow a mass-size relation unlike ``normal" GCs (see Fig. \ref{fig:1}). 
The second is associated with the stripping of galaxy nuclei by larger galaxies, yielding 
objects which display a similar mass-size relation. This study will be described in detail 
in a forthcoming paper \cite{Norris&Kannappan}.

\section{Results}
\label{sec:1}

\begin{figure}
\resizebox{0.975\columnwidth}{!}{%
  \includegraphics{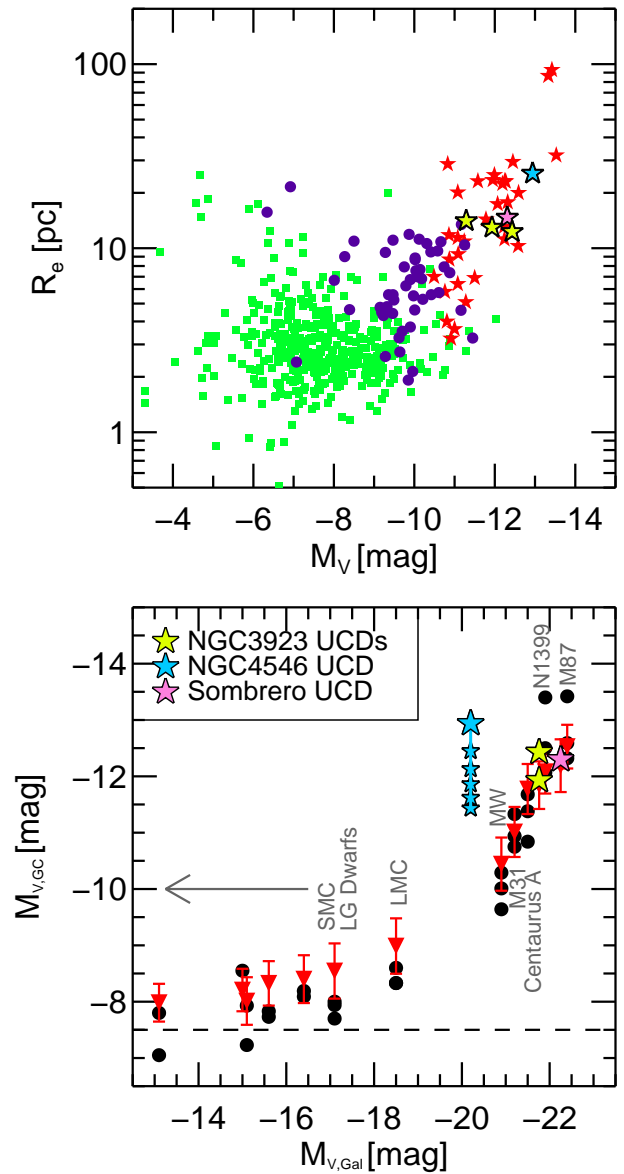} }
\caption{
\textbf{Upper~Panel:~}
Half-light radius vs. absolute V-band magnitude for compact stellar systems. 
Green squares are Milky Way and M31 GCs, purple circles are Cen A GCs,
red stars are Virgo and Fornax UCDs (sources are provided in full in
\cite{Norris&Kannappan}).
\textbf{Lower~Panel:~} The absolute magnitude of the brightest two or three GCs/UCDs of 
a galaxy as a function of host galaxy total luminosity (after Figure 2 of \cite{Hilker09}). 
Data come from the compilation of \cite{Hilker09} plus NGC3923, NGC4546 
and the Sombrero. The smaller blue stars are predictions for the future 
evolution of NGC4546 UCD1 for ages (top to bottom) 5, 7, 9, 11 and 13~Gyr.
The red triangles with error bars indicate the average luminosity of the 
brightest GC found in 10,000 Monte-Carlo simulations of the GCLF of each 
galaxy, assuming the measured total number of clusters, the universal GCLF 
turnover, and the measured dispersion of the GCLF.}
\label{fig:1}       
\end{figure}

We find that using multiple lines of evidence it is possible to robustly classify 
UCDs as either GGCs or stripped-nucleus UCDs. 
One powerful diagnostic of UCD origin, first presented by \cite{Hilker09}, is a 
plot of the magnitude of the brightest GC/UCD of a galaxy versus either (i) galaxy 
total luminosity or (ii) the total number of GCs associated 
with the host galaxy (see Fig.~\ref{fig:1}). This plot allows one to probe whether 
a particular UCD is statistically consistent with being a highly luminous member of 
the host galaxy's GC system. Interestingly, the UCD of NGC4546, which we suspect
for other reasons given below to be a stripped nucleus, is an extreme outlier 
in this plot, as are the most luminous UCDs of Virgo and Fornax. 
Other clues to the true nature of a UCD are provided by its age and [$\alpha$/Fe]
(especially when compared to those of the GCs of its host galaxy), its expected
dynamical friction decay timescale based on projected distance from the host,
and any indications of past merger events likely to have involved the stripping 
of nuclei (such as tidal tails, counterrotating gas in the host galaxy, etc).\\

\noindent Using these clues we find that the UCDs of NGC3923 are almost certainly 
GGCs:
\begin{itemize}
	\item NGC3923 has multiple UCDs (at least 2, with 1 more highly likely).
	\item As in Cen A (see Fig.~\ref{fig:1} and \cite{Taylor10}) several 
	NGC3923 objects previously classified as GCs smoothly transition 
	between GC-like and UCD-like mass-size behaviour.
	\item The UCDs smoothly extend the colour-magnitude relation of the blue GCs.
	\item All three possible UCDs have long apparent dynamical friction decay 
	timescales ($>$ 7 Gyr).
	\item The UCDs are all consistent with the bright end of the GC 
	luminosity function (GCLF, see Fig.~\ref{fig:1}).
\end{itemize}	
	
\noindent The UCD of NGC4546 is almost certainly a stripped nucleus:
\begin{itemize}	
	\item This UCD is young ($\sim$3~Gyr old), whereas NGC4546 has a uniformly old 
	($\sim$10~Gyr) stellar population.
	\item It does not seem to be associated with an equivalently young 
	GC population.
	\item Its apparent dynamical friction decay timescale is short ($\sim$0.5~Gyr).
	\item It is not explainable as the bright extension of the GCLF of NGC4546
	(see Fig.~\ref{fig:1}). Even when the UCD has aged to 13~Gyr it
	will be significantly overluminous relative to the expectations of a model
	where it is the most luminous member of NGC4546's GC system.
	\item Both the UCD and the gas of NGC4546 counterrotate relative to the stellar 
	disk, suggesting a common external origin for them both.
\end{itemize}	

\noindent The Sombrero UCD is also a probable but uncertain GGC:
\begin{itemize}	
	\item Its stellar population properties (Age$\sim$12.6~Gyr, [Fe/H] = -0.08, and 
	[$\alpha$/Fe] = 0.06) are consistent with those of the GC system 
	of the Sombrero \cite{Larsen02} but may also be consistent with an
	ancient stripping event.
	\item It has a long apparent dynamical friction decay timescale 
	(greater than a Hubble time).
	\item It is statistically consistent with an extension of the GCLF (see Fig.~\ref{fig:1}),
	but no intermediate objects are known as in NGC3923.
\end{itemize}

Despite the compelling evidence for two different UCD formation channels
presented by NGC3923 and NGC4546, we find that both UCD types have 
very similar colour-magnitude and mass-size diagrams. 
In fact, we find that the loci of blue/red GCs, blue/red UCDs and dwarf/giant 
galaxy nuclei, respectively, are indistinguishable in a colour-magnitude diagram,
with dwarf nuclei and blue UCDs obeying the same mass-metallicity (``blue tilt")
trend as massive blue GCs.
This observation indicates that at fainter magnitudes objects traditionally 
classified as GCs may comprise a composite population
of ``normal" GCs and stripped nuclei too small to qualify as UCDs.

\section{Conclusions}

In a study of UCDs in field/group environments we have found unambiguous 
evidence of two UCD formation channels: stripped-nucleus and giant GC (GGC).
The UCD of NGC4546 provides the first clear example of a young 
UCD formed in a minor merger event by the stripping of a companion galaxy 
(complementing the possible young GGC described by \cite{Maraston04}).
The multiple UCDs of NGC3923 and the single UCD of the Sombrero are likely 
to be GGCs, with properties matching their respective GC systems. As GCs
are generally thought to form during the major mergers responsible for spheroid
formation, we therefore suggest that UCDs can provide a useful probe of 
both the major and minor merger histories of galaxies.


\end{document}